\documentclass[12pt]{JHEP3}
\usepackage{mathrsfs}
\usepackage{amsmath,amssymb}
\usepackage{epsfig}
\usepackage{relsize}
\usepackage{graphicx}

\input epsf

\def\x'{\mathaccent 19 x}
\def\y'{\mathaccent 19 y}
\def\n'{\mathaccent 19 n}
\def\u'{\mathaccent 19 u}

\def\et'{\mathaccent 19 \eta}
\def\th'{\mathaccent 19 \theta}
\def\lam'{\mathaccent 19 \lambda}
\def\varet'{\mathaccent 19 \vartheta}
\def\rh'{\mathaccent 19 \rho}
\def\ph'{\mathaccent 19 \phi}
\def\xb'{\mathaccent 19 {\bar{x}}}

\def\sl(2){\alg{sl}(2)}

\def\be{\begin{equation}}
\def\ee{\end{equation}}

\newcommand{\bea}{\begin{eqnarray}}
\newcommand{\eea}{\end{eqnarray}}

\newcommand{\alg}[1]{\mathfrak{#1}}
\newcommand{\su}{\alg{su}}

\newcommand{\AdS}{{\rm  AdS}_5\times {\rm S}^5}

\newcommand{\sfrac}[2]{{\textstyle\frac{#1}{#2}}}

\newcommand{\bem}{\left (\begin{matrix}}
\newcommand{\eem}{\end{matrix} \right )}

\newcommand{\beq}{\begin{equation}}
\newcommand{\eeq}{\end{equation}}
\newcommand{\beqa}{\begin{eqnarray}}
\newcommand{\eeqa}{\end{eqnarray}}
\newcommand{\beqar}{\begin{eqnarray*}}
\newcommand{\eeqar}{\end{eqnarray*}}
\numberwithin{equation}{section}

\author{L.~F. Alday$^a$\footnote{Email:
l.f.alday, g.arutyunov, b.eden@phys.uu.nl}\, , G.
Arutyunov$^a$\footnote{Correspondent fellow at Steklov
Mathematical Institute, Moscow.},\,  M.~K. Benna$^b$
\footnote{Email: mbenna, klebanov@princeton.edu},\, B. Eden$^a$,\,
I.~R.
 Klebanov$^b$
\\
$^{a}$ {\it
Institute for Theoretical Physics and Spinoza Institute,\\
Utrecht University, 3508 TD Utrecht, The Netherlands} \\
$^{b}$ {\it Department of Physics and
Princeton Center for Theoretical Physics,\\ Princeton University,
 Princeton, NJ  08544} }

\abstract{We give an exact analytic solution of the strong
coupling limit of the integral equation which was recently
proposed to describe the universal scaling function of high spin
operators in ${\cal N}=4$ gauge theory. The solution  agrees with
the prediction from string theory, confirms the earlier numerical
analysis and provides a basis for developing a systematic
perturbation theory around strong coupling.}

\title{On the Strong Coupling Scaling Dimension \\
of High Spin Operators }

\preprint{\smaller{\smaller{\smaller{ITP-UU-07-06}}}\\[-.5ex]
          \smaller{\smaller{\smaller{PUPT-2222}}}\\[-.5ex]
          \smaller{\smaller{\smaller{SPIN-07-06}}} }

\begin{document}

\section{Introduction}
The scaling dimension $\Delta$ for twist two operators at large
values of the Lorentz spin $S$ is characterized by the universal
scaling function (cusp anomalous dimension) $f(g)$:
$$\Delta-S=f(g) \ln S+\dots ,$$
with $g=\sqrt{g_{YM}^2 N}/4\pi$. The logarithmic dependence of the
dimension on large Lorentz spin is a generic feature that has been
independently observed for both ${\cal N}=4$ SYM and the
corresponding string theory dual, see, e.g.,
\cite{Kotikov:2001sc}-\cite{Frolov:2002av}. Importantly, such a
scaling behavior stems \cite{Belitsky:2006en,Eden:2006rx} from the
large spin limit of the Bethe equations
\cite{Beisert:2004hm}-\cite{Beisert:2005fw} which underlie the
integrable structures of gauge and string theories. A linear
integral equation  arising in the large spin limit from the
postulated set of gauge/string Bethe equations has been recently
derived in \cite{Beisert:2006ez}. We will refer to it as the BES
equation. This equation allows one to compute the universal scaling
function $f(g)$ to any desired order in perturbation theory. For
instance, for the first few orders one obtains
\begin{equation}
\nonumber f(g)=8 g^2 -\frac{8}{3}\pi^2 g^4+\ldots
\end{equation}
Remarkably, independent gauge theory computations
\cite{Bern:2006ew,{Cachazo:2006az}} confirm this result up to four
loop orders, {\it i.e.} up to $g^8$.

\medskip

Schematically,  the BES equation has the following form
$$\sigma(t)=1+g^2\int K(t,t') \sigma(t') dt'\, ,$$
where $K(t,t')=K^{(m)}(t,t')+\kappa K^{(c)}(t,t')$ and $\kappa=2$.
Here $K^{(m)}$ is the so-called main scattering kernel and
$K^{(c)}$ is the dressing kernel. The dressing kernel arises upon
taking into account an additional scattering phase in the S-matrix
 which is apparently needed to achieve an
agreement with string theory predictions in the strong coupling
limit \cite{Arutyunov:2004vx}. This phase is also the only freedom
in the S-matrix which cannot be fixed by symmetry arguments
\cite{Beisert:2005tm}.\footnote{See
\cite{Hofman:2006xt}-\cite{Arutyunov:2006yd} for recent studies of
the string S-matrix and its symmetries.} The form of the phase is
largely but not completely constrained by an additional
requirement of crossing symmetry \cite{Janik:2006dc}. The
constraints arising from crossing symmetry were shown to hold
\cite{Arutyunov:2006iu} at two leading orders
 in string perturbation
theory \cite{Arutyunov:2004vx,Hernandez:2006tk}. Recently an
interesting solution for the crossing symmetric phase has been
obtained \cite{Beisert:2006ib} which agrees nicely with all
available string and field-theoretic data.

\smallskip

The agreement observed at the four loop level between the scaling
function of BES and the results obtained in the field-theoretic
framework is very important, because it implies the appearance of
the additional scattering phase in perturbation theory and,
therefore, it distinguishes the BES equation from the earlier
proposal \cite{Eden:2006rx} which corresponds to taking
$\kappa=0$.

\smallskip

According to the AdS/CFT duality \cite{Maldacena:1997re} the large
$g$ behavior of $f(g)$ can be extracted by studying the energy of
the folded string spinning in the ${\rm AdS}_3$ part of the target
space (the GKP solution) and it turns out to be
\cite{Gubser:2002tv,{Frolov:2002av}}
\begin{equation}
\label{largeg}
f(g)=4 g-\frac{3 \ln 2}{\pi}+...
\end{equation}
In the recent work \cite{Benna:2006nd} and \cite{Kotikov:2006ts}
an important question was addressed, namely, how the perturbative
expansion controlled by the BES equation can be extrapolated to
large values of the coupling constant and wether the resulting
expression is consistent with the string theory predictions. In
particular, in \cite{Benna:2006nd} a numerical method was devised
to solve the BES equation around the strong coupling point and
numerical solutions were shown to be in perfect agreement with
eq.~(\ref{largeg}), providing a non-trivial test of the AdS/CFT
correspondence.

\smallskip

In this note we find an exact analytic expression for the density
$\sigma(t)$ in the strong coupling limit. If one insists on
truncating the BES equation at leading order for large $g$ the kernel
becomes degenerate and therefore a unique solution for $\sigma(t)$
cannot be found without additional assumptions. However, for
finite values of $g$ the solution is unique and can be found
numerically with high precision. These two statements are consistent because expanding the BES equation in a power series in $1/g$ we find a second equation at subleading order that removes any ambiguity in the strong coupling solution $\sigma(t)$.

\smallskip

To gain some intuition our approach will be to first
investigate the solution numerically for finite values of the
coupling, which will suggest a simple additional requirement that
should be imposed on the underdetermined, truncated BES equation to single out a unique solution in the strong
coupling limit which is the convergence point of our numerics.
After this additional requirement is identified, we find an analytic solution for $\sigma(t)$ in the strong coupling
limit. We then show that this solution is in fact completely determined by taking into account the subleading contributions to the BES equation,
thus confirming analytically the validity of the auxiliary condition suggested by the numerics and of our strong coupling solution $\sigma(t)$.

\smallskip

Upon making the Fourier transform to the rapidity $u$-plane
the corresponding density $\sigma(u)$ reads
$$
\sigma(u)=\frac{1}{4\pi g^2
}\left(1-\frac{\theta(|u|-1)}{\sqrt{2}}\sqrt{1+\frac{1}{\sqrt{1-\frac{1}{u^2}}}}\right)\,
,
$$
where $\theta(u)$ is the step function. Thus, on the rapidity plane
the leading density is an algebraic function which is constant in
the interval $|u|<1$. We see that, in contrast to the weak coupling
solution of the BES equation \cite{Beisert:2006ez}, the strong
coupling density exhibits a gap
between $[-\infty,-1]$ and $[1,\infty]$, see Fig.1. Remarkably, this
is reminiscent of the behavior of the corresponding solutions
describing classical spinning strings.
\vskip 10pt \noindent
\begin{minipage}{\textwidth}
\begin{center}
\includegraphics[width=0.50\textwidth]{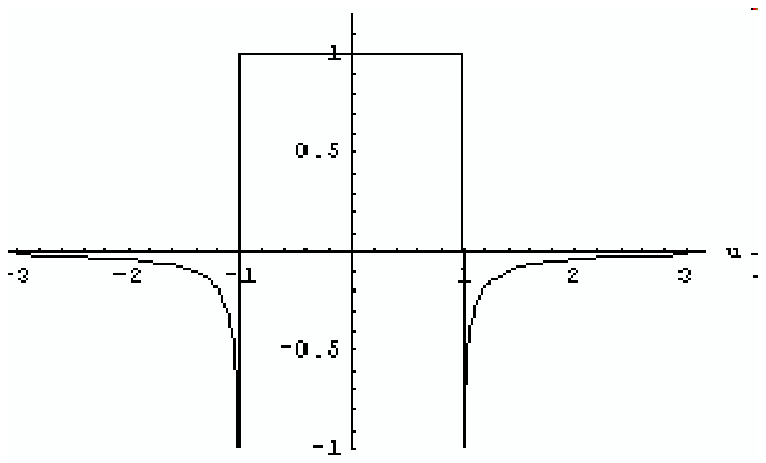}
\end{center}
\end{minipage}
\vskip 15pt
\begin{center}
\parbox{5in}{{\footnotesize Fig.1.
The plot of the analytic solution for the leading density $4\pi
g^2\sigma(u) $. } }
\end{center}
We also would like to point out that the strong coupling solution of
the BES equation we have found is apparently different from the
singular solution $\chi_{sing}$ discussed in \cite{Kotikov:2006ts}.

\smallskip

The paper is organized as follows. In section 2 we analyze the
matrix form of the BES equation at strong coupling, first
numerically and then analytically. We compute the coefficients in
the expansion of $\sigma(t)$ in terms of Bessel functions and
based on this numerical analysis we make a guess of how these
coefficients could be (partially) related to each other at strong
coupling. In the strong coupling limit we obtain an analytic
equation for the coefficients which turns out to have a degenerate
kernel. Supplementing this equation with the proposed relation
among the coefficients allows us to find an exact analytical
solution for $\sigma(t)$. We then show that in fact no guess is
necessary if one expands the BES equation to higher order in
$1/g$, which leads to a second equation that singles out our
solution as the correct one.

\smallskip

We proceed to
find an analogous pair of equations for the subleading coefficients at strong
coupling. Again they appear at different orders in the inverse coupling constant,
each of them individually being a degenerate, half-rank equation.
We identify some constraints on the subleading
value of $\sigma(t)$, but postpone a detailed
analysis of the strong coupling perturbation theory to a future publication.
In section 3
we analyze the BES equation at strong coupling in the Fourier
space and verify that the results of section 2 agree with this
analysis. Finally, we end with conclusions and open
problems.

\section{BES matrix equation at strong coupling}
In this section we study the density of fluctuations $\sigma(t)$
for large values of the coupling constant. First we argue, by
performing numerical analysis, that $\sigma(t)$ obeys a certain
additional requirement. Then we consider the matrix BES equation
at leading order in the large $g$ expansion, use the previously
found requirement to solve for the density and confirm by expanding
the BES equation further that this solution is in fact complete determined analytically.
Finally, we derive and briefly discuss constraints on the subleading value of $\sigma(t)$.

\subsection{Numerical studies of $\sigma(t)$}
The universal scaling function is related to $\sigma(t)$ as
follows
\begin{equation}
f(g)=16 g^2 \sigma(0) \, ,
\end{equation}
where the density of fluctuations $\sigma(t)$ satisfies the
following integral equation \cite{Eden:2006rx,Beisert:2006ez}

\begin{equation}
\label{BES} \sigma(t)=\frac{t}{e^t-1}\Big( K(2 g t,0)-4 g^2
\int_0^\infty {\rm d}t'\,  K(2 g t,2 g t')\,\sigma(t') \Big)\, .
\end{equation}
The kernel separates  into two pieces, the main scattering kernel
$K^{(m)}$ and the dressing kernel $K^{(c)}$ \bea
K(t,t')=K^{(m)}(t,t')+2 K^{(c)}(t,t')\, , \eea which are expressed
in terms of Bessel functions as follows \bea \nonumber
K^{(m)}(t,t')&=&\frac{J_1(t)J_0(t')-J_0(t)J_1(t')}{t-t'}\, ,\\
\nonumber K^{(c)}(t,t')&=&4g^2 \int_0^\infty {\rm d}t''\,
K_1(t,2gt'') \frac{t''}{e^{t''}-1}K_0(2gt'',t') \, .\eea Here
$K_0$ and $K_1$ denote even and odd parts of the kernel under
change of the sign of the arguments:
\begin{eqnarray}
\nonumber K_0(t,t')&=&\frac{t
J_1(t)J_0(t')-t'J_0(t)J_1(t')}{t^2-t'^2}=\frac{2}{tt'}\sum_{n=1}^\infty
(2n-1)J_{2n-1}(t)J_{2n-1}(t')\, ,\\
\label{defK01} K_1(t,t')&=&\frac{t'
J_1(t)J_0(t')-tJ_0(t)J_1(t')}{t^2-t'^2}=\frac{2}{tt'}\sum_{n=1}^\infty
2nJ_{2n}(t)J_{2n}(t') \, .
\end{eqnarray}
As was shown in \cite{Benna:2006nd}, if one introduces a function
\begin{equation}
s(t)=\frac{e^t-1}{t}\sigma(t)\,
\end{equation}
and expands $s(t)$ in the following series
\begin{equation}\label{basis}
s(t)=\sum_{n=1}^\infty s_n \frac{J_n(2 g t)}{2gt}
\end{equation}
then eq.(\ref{BES}) transforms into a matrix equation for the
vector $s=(s_1,s_2,s_3,...)^t$
\begin{equation}
\label{matrixeq} (1+K^{(m)}+2 K^{(c)}) s=(1+2 C)e \, .
\end{equation}
Here $e=(1,0,0,0,...)^t$ and $K^{(m)}$, $K^{(c)}$, $C$ are
infinite-dimensional matrices with the entries \bea
K^{(m)}_{mn}&=&2(N Z)_{mn}\, , \nonumber \\
K^{(c)}_{mn}&=&2(C Z)_{mn}\, ,\nonumber \\
C_{mn}&=&2(P N Z Q N)_{mn}\, , \nonumber \eea where $Q={\rm
diag}(1,0,1,0,...)$, $P={\rm diag}(0,1,0,1,...)$ and $N={\rm
diag}(1,2,3,...)$. The coupling constant enters into the matrix
$Z$ whose entries are given by
\begin{equation}
\label{Zmn} Z_{mn} = \int_0^\infty {\rm d}t\, \frac{J_m(2 g t)
J_n(2g t)}{t(e^t-1)}\, .
\end{equation}

The key observation is that for intermediate
values of $g$, for instance $2<g<20$, one can approximate the
infinite-dimensional matrices entering the BES equation by
matrices of finite rank $d$, with $d$ not much larger than $g$.
With matrices of finite rank it is possible to solve numerically
for the coefficients $s_k$ for different values of the coupling
constant and to find the best fit result for an expansion of the
type
\begin{equation}
\label{lsc} s_k=\frac{1}{g}s_k^{\ell}+\frac{1}{g^2}s_k^{s\ell}+\dots
\end{equation}
As the numerical analysis indicates, the finite rank approximation
is valid for computing the coefficients $s_k$ with $k \ll d$. We
have solved numerically eq.(\ref{matrixeq}) for numerous points in the
range $2<g<20$ and use $d=50$. In the table below  the values for
a few leading coefficients $s_k^{\ell}$ are exhibited.
\begin{equation}
\nonumber
\begin{tabular}{|c|c|c|c|}
  \hline
  $k$ & $s^{\ell}_{2k-1}$ & $s^{\ell}_{2k}$ & $100|(s^{\ell}_{2k-1}-s^{\ell}_{2k})/s^{\ell}_{2k}|$\\
  \hline
  1 & 0.500006 & 0.499993 & 0.003\\
  2 & -0.75005 & -0.74977 & 0.038\\
  3 & 0.93727 & 0.93676 & 0.055\\
  4 & -1.09281 & -1.09415 & 0.12\\
  5 & 1.2239 & 1.2333 & 0.77\\ \hline
\end{tabular}
\end{equation}
Some comments are in order. First, notice that the value for
$s_1^{\ell}$ is in perfect agreement with the value predicted from
string theory $s_1^{\ell}=1/2$ and confirmed numerically (with a
precision higher than the one presented here) by
\cite{Benna:2006nd}. Second, notice that the difference between
$s_{2k-1}^{\ell}$ and $s_{2k}^{\ell}$ is in all the cases smaller
than $1\%$, and gets bigger as $k$ increases; this is related to
the fact that the rank of the matrices is finite.

Thus, the numerical analysis suggests that in the limit of
infinite rank matrices the following relation holds for the
leading coefficients in the strong coupling expansion
\begin{equation}
\label{leadingcond} s_{2k-1}^{\ell}=s_{2k}^{\ell}\, .
\end{equation}
As we will see later on, this condition will allow one to solve
analytically for the coefficients $s_k^{\ell}$ and the values
obtained will be in perfect agreement with the ones computed
numerically.

Further evidence comes from the fact that one can approximate the
matrix elements $Z_{mn}$ by their analytic values at strong
coupling (see next subsection). Therefore, one can consider
matrices of much higher rank, fix a sufficiently large value of
$g$ and compute (numerically) the coefficients $s_k^{\ell}$. Below
we present the results for $g=10000$ and $d=250$.

\begin{equation}
\nonumber
\begin{tabular}{|c|c|c|c|}
  \hline
  $k$ & $s^{\ell}_{2k-1}$ & $s^{\ell}_{2k}$ & $100|(s^{\ell}_{2k-1}-s^{\ell}_{2k})/s^{\ell}_{2k}|$\\
  \hline
  1 & 0.49993 & 0.49991 & 0.0049\\
  2 & -0.74943 & -0.74938 & 0.0073\\
  3 & 0.93585 & 0.93577 & 0.0085\\
  4 & -1.09033 & -1.09023 & 0.0089\\
  5 & 1.22455 & 1.22444 & 0.0087\\ \hline
\end{tabular}
\end{equation}
\vskip 0.5cm

\noindent As $d$ increases, we see that the difference between
$s_{2k-1}^{\ell}$ and  $s_{2k}^{\ell}$ (for $k=2,3,4$) decreases
considerably. Also, the difference is approximately constant for
small values of $k$. We should stress, however, that a priory
there is no reason to expect the results obtained by keeping in
the BES equation only the leading term for $Z$ to be valid, since as we will see
the subleading terms in the matrix elements $Z_{mn}$ are
necessary to fix uniquely the leading order solution for $s_k$.
Surprisingly, one still obtains a good approximation to the large $g$ solution in this fashion. Nevertheless, we regard the
present computation as less robust.

To conclude, requiring continuity in $g$, the leading coefficients
$s_k^{\ell}$ in the strong coupling expansion of the function
$\sigma(t)$ exhibit the relation (\ref{leadingcond}), which
constraints a possible form of $\sigma(t)$.

\subsection{Analytic solution at strong coupling}
For finite real values of $g$ the matrix element $Z_{mn}$ is given
by a convergent integral. However, it is not obvious if it is
possible to express the result of integration as a power series in $1/g$. Indeed, expanding the integrand in eq.(\ref{Zmn}) as
\begin{equation}
\nonumber Z_{mn}=\int_0^\infty {\rm d}t\, \frac{J_m(t)
J_n(t)}{t}\left( \frac{2g}{t}-\frac{1}{2}+\frac{t}{24g} +\ldots
\right)\, ,
\end{equation}
leads to a power series (with coefficients given essentially by
the Bernoulli numbers) which converges only for $|t/(2g)| < 2\pi$.
We see that only the first two leading terms in this expansion can
be integrated, while already in the third term divergent integrals
appear. It is therefore natural to assume that the expansion of
$Z$ develops as \bea Z = g Z^{\ell}+Z^{s\ell}+\ldots
,\label{suspexp}\eea where the first two terms, $Z^{\ell}$ and
$Z^{s\ell}$,  are given by the convergent integrals mentioned
above. Explicitly, they are \bea \nonumber
Z^{\ell}_{mn}&=&-\frac{8}{\pi}\,
\frac{\cos((m-n)\pi/2)}{(m+n+1)(m+n-1)(m-n+1)(m-n-1)}\, ,\\
\label{Zls}
Z^{s\ell}_{mn}&=&-\frac{1}{\pi}\frac{\sin((m-n)\pi/2)}{m^2-n^2}\,
. \eea Assumption (\ref{suspexp}) is well supported by the
numerics.
We have checked that numerical values for $Z_{mn}$ for large $g$ are
in a very good agreement with the analytic expressions (\ref{Zls}).
On the other hand, the status of the higher order terms in
eq.(\ref{suspexp}) is not clear to us\footnote{For instance, whether
expansion (\ref{suspexp}) contains terms of the type $\log(g)/g$,
etc. From the string theory point of view, the appearance of such
terms could be an indication of a two-loop divergence of the
world-sheet S-matrix. Further analysis is needed to clarify this
issue.}. Thus, in what follows, we largely restrict our
investigation of the BES equation to the first two leading terms.

\medskip

With this word of caution we proceed to investigate eq.
(\ref{matrixeq}) in the strong coupling limit. It is not hard to see
that it implies the following equation for the leading vector
$s^{\ell}$
\begin{equation}
\label{leadingeq} K^{(c)\, \ell} s^{\ell}=C^{\ell} e \, ,
\end{equation}
where the leading matrices $K^{(c)\, \ell}$ and $C^{\ell}$ are
obtained by keeping the leading contribution $Z^{\ell}$ only. It
turns out that  for even values of $d$ the kernel $K^{(c)\, \ell}$
has rank $d/2$, hence from the equation above it is possible to
solve only for half of the components of $s$. However, now the
condition (\ref{leadingcond}) enters into play. Imposing
eq.(\ref{leadingcond}) together with eq.(\ref{leadingeq}) allows
one to uniquely determine the vector $s^{\ell}$.

\medskip In order to find the solution in the limit of infinite $d$
it is convenient to express the leading equation in the following
way

\begin{equation}
\label{evenodd} K^o s^o+K^e s^e=\frac{1}{2} e\, ,
\end{equation}
where $s^o,~s^e$ are vectors of length $d/2$ comprising the odd
and even components of $s^{\ell}$ respectively:
$s^o=(s_1^{\ell},s_3^{\ell},s_5^{\ell},...)^t$ and
$s^e=(s_2^{\ell},s_4^{\ell},s_6^{\ell},...)^t$. Further
$e=(1,0,0,...)^t$. It is easy to check that eqs.(\ref{evenodd}) and
(\ref{leadingeq}) are equivalent provided
\begin{equation}
(K^o)_{mn}=Z^{\ell}_{2m-1,2n-1}+Z^{\ell}_{2m+1,2n-1}\, , ~~~~~
(K^e)_{mn}=Z^{\ell}_{2m-1,2n}+Z^{\ell}_{2m+1,2n}\, .
\end{equation}
Then, we find that the unique solution of eq.(\ref{evenodd})
satisfying the relation $s^o=s^e$ turns out to be
\begin{equation}
\label{sol}
s_{2k-1}^{\ell}=s_{2k}^{\ell}=(-1)^{k+1}\frac{\Gamma(k+\sfrac{1}{2})}{\Gamma(k)\Gamma(\sfrac{1}{2})}\,
.
\end{equation}
This remarkably simple expression for the coefficients
$s^{\ell}_k$ is the main result of this paper. By using the
following identities (true in the limit of infinite rank)
\begin{eqnarray}
\nonumber
(K^{e~-1})_{mn}&=&-4(-1)^{m+n+1} m n^2, \hspace{0.2in} n \le m\\
&=& -4(-1)^{m+n+1} m^3, \hspace{0.2in}~~ n > m  \label{inverses} \\
(K^{e~-1}
K^o)_{mn}&=&(-1)^{m-n}\frac{32}{\pi}\frac{m^3}{(4m^2-(1-2n)^2)(1-2n)^2}
\nonumber
\end{eqnarray}
 one can check explicitly that the coefficients
(\ref{sol}) indeed solve eq.(\ref{evenodd}). Thus, we found that
restricting the coefficients $s_k$ to the leading order
expressions $s_k^{\ell}$ the function $s(t)$ is given by \bea
s(t)=\frac{1}{g}\sum_{k=1}^{\infty}(-1)^{k+1}\frac{\Gamma(k+\sfrac{1}{2})}{\Gamma(k)\Gamma(\sfrac{1}{2})}
\frac{J_{2k}(2gt)+J_{2k-1}(2gt)}{2gt}\, ,\label{sexp}\eea where we
have omitted the subleading contributions. The last formula can be
considered as the leading term in the large $g$ expansion of the
density $s(t,g)$ with $gt$ kept finite. As is clear from
eq.(\ref{BES}), we are only interested in values of $s(t)$ for
$t\geq 0$. The series can be summed and for this range of $t$ the
result expressed in terms of the confluent hypergeometric function
of the second kind $U(a,b,x)$:
\begin{eqnarray}
\nonumber s(t) & = & - \frac{i}{8 \pi g^2 t} \, e^{2i g t
\phantom{-}} \Big(\Gamma(\sfrac{3}{4}) \, U(-\sfrac{1}{4},0,- 4 i
g t) \, + \, \Gamma(\sfrac{5}{4})\,
U(\sfrac{1}{4},0,-4 i g t) \Big) \\
& & + \frac{i}{8 \pi g^2 t} \, e^{- 2 i g t }
\Big(\Gamma(\sfrac{3}{4}) \, U(-\sfrac{1}{4},0, 4 i g t ) \, + \,
\Gamma(\sfrac{5}{4}) \, U(\sfrac{1}{4},0,4 i g t) \Big)\, .
\label{Fprof}
\end{eqnarray}
The leading density (\ref{Fprof}) has a rather complicated
profile. For $gt\to 0$ it perfectly reproduces the desired result
$s(t)\to 1/4g$. On the other hand, the asymptotic of $s(t)$ for
$gt\to \infty $ exhibits a highly oscillating behavior.

\subsection{An alternative derivation of the strong coupling solution} \label{lderiv}

Let us now show that the result (\ref{sol}) can in fact be derived without resorting to any auxiliary conditions obtained from numerical arguments. The degenerate equation appearing at leading order, which can be used to express one half of the coefficients $s_n$ in terms of the other, can be supplemented by another half-rank equation from subleading terms in the BES equation. Together they determine a unique solution.

\smallskip
Writing out the integral equation (\ref{BES}) in the basis (\ref{basis}) with explicit matrix indices we find
\begin{eqnarray}\label{BES2}
s_n {J_n(2gt)\over 2gt} &=&  {J_1(2gt)\over 2gt} + 8n Z_{2n,1} {J_{2n}(2gt)\over 2gt} -  2 n Z_{nm} s_m  {J_n(2gt)\over 2gt}  \nonumber\\ &&- 16 n(2m-1)  Z_{2n,2m-1}Z_{2m-1,r} s_r  {J_{2n}(2gt)\over 2gt}\ ,
\end{eqnarray}
where all indices are summed over from $1$ to $\infty$. Now we
split up the integral equation according to powers of $g$ and into
odd and even rows (indices of Bessel functions).

\smallskip
At $\mathcal{O}(g)$ the odd equation is trivial and the even one reads:
\begin{equation} \label{O0e}
2(2m-1)Z^{\ell}_{2n,2m-1}Z^{\ell}_{2m-1,r} s^{\ell}_r = Z^{\ell}_{2n,1} = {1 \over 4} \delta_{n,1}\ .
\end{equation}
This is precisely eq.(\ref{leadingeq}) employed in the previous subsection. At $\mathcal{O}(1)$ the odd rows lead to the condition
\begin{equation} \label{O1o}
Z^{\ell}_{2m-1,r} s^{\ell}_r = {1 \over 2} \delta_{m,1}\ .
\end{equation}
Actually this equation implies the previous one. It determines one half of the coefficients $s^{\ell}_n$ in terms of the other. Expanding further, at $\mathcal{O}(1)$ the even equation is given by
\begin{eqnarray} \label{O1e}
&&8n Z^{s\ell}_{2n,1} - 4nZ^{\ell}_{2n,m} s^{\ell}_m - 16n(2m-1)Z^{\ell}_{2n,2m-1}Z^{\ell}_{2m-1,r} s^{s\ell}_r  \nonumber \\ &-&  16n(2m-1)Z^{\ell}_{2n,2m-1}Z^{s\ell}_{2m-1,r} s^{\ell}_r - 16n(2m-1)Z^{s\ell}_{2n,2m-1}Z^{\ell}_{2m-1,r} s^{\ell}_r= 0\ .\qquad
\end{eqnarray}
To determine the other half of the coefficients $s^{\ell}_n$ we need to eliminate $s^{s\ell}_n$ from this equation. To do this we examine the odd equation at $\mathcal{O}(1/g)$
\begin{equation} \label{O2o}
- 2(2m-1)Z^{\ell}_{2m-1,r} s^{s\ell}_r - 2(2m-1)Z^{s\ell}_{2m-1,r} s^{\ell}_r = s^{\ell}_{2m-1}\ .
\end{equation}
Now we use (\ref{O1o}) and (\ref{O2o}) to simplify (\ref{O1e}), which gives
\begin{eqnarray}
Z^{\ell}_{2n,m} s^{\ell}_m  - 2 Z^{\ell}_{2n,2m-1} s^{\ell}_{2m-1}  = Z^{\ell}_{2n,m} (-1)^m s^{\ell}_m =  0\ .
\end{eqnarray}
This is the second equation we were looking for, which together with (\ref{O1o}) completely determines $s^{\ell}_n$. If we define a matrix $\tilde{Z}_{nm}$ which is identical to  $Z_{nm}$ except for a sign flip when both $n$ is even and $m$ is odd, we can combine the two conditions into the full-rank equation for the strong coupling solution
\begin{eqnarray} \label{lfullrank}
\tilde{Z}^{\ell}_{nm} s^{\ell}_m =  {1 \over2} \delta_{n,1}\ .
\end{eqnarray}
Note that the additional minus signs in the definition of $\tilde{Z}_{nm}$ arise precisely because of the introduction of the dressing kernel, and would by absent for $\kappa=0$.
Writing out (\ref{lfullrank}) explicitly shows that the $s^{\ell}_n$ have to satisfy
\begin{eqnarray} \label{sum1}
\sum_{k=1}^\infty { 4(-1)^{n+k+1} s^{\ell}_{2k-1} \over (2n-2k-1)(2n-2k+1)(2n+2k-3)(2n+2k-1)\pi} \qquad\qquad\nonumber\\   = - {s^{\ell}_{2n}\over 8n(2n-1)} - {s^{\ell}_{2n-2}\over 8(n-1)(2n-1)}  + {1 \over4} \delta_{n,1}\ , \\ \label{sum2}
\sum_{k=1}^\infty { 4(-1)^{n+k+1} s^{\ell}_{2k} \over (2n-2k-1)(2n-2k+1)(2n+2k-1)(2n+2k+1)\pi} \qquad\qquad\nonumber\\ = {s^{\ell}_{2n+1}\over 8n(2n+1)} + {s^{\ell}_{2n-1}\over 8n(2n-1)}\ ,
\end{eqnarray}
for all $n \ge 1$ (where it is understood that the second term on the right hand side of (\ref{sum1}) is absent for $n=1$).
Indeed these equations are obeyed by
\begin{eqnarray} \label{slsol}
s^{\ell}_{2n-1} =  s^{\ell}_{2n} = {(-1)^{n-1}(2n-1)!! \over 2^n (n-1)!}\ ,
\end{eqnarray}
which is precisely the solution (\ref{sol}) found in the previous
subsection. To show this, note that the coefficients
$s^{\ell}_{2n-1}$ are generated by the Taylor expansion of
$(1+x)^{-3/2}$, which makes it easy to perform the above sums as
integrals over functions of the form $x^m (1-x^2)^{-3/2}$ for some
appropriate power $m$ chosen to generate the necessary terms in
the denominators of the left hand sides of eqs.(\ref{sum1}) and
(\ref{sum2}).

\subsection{Fluctuation density in the rapidity plane}

To get more insight into the structure of the leading solution, we
find it convenient to perform the (inverse) Fourier transform of
the density $\sigma(t)\to \sigma(u)$:
\begin{equation}
\label{ft} \sigma(u)=\frac{1}{2\pi}\int_{-\infty}^\infty {\rm d}t~
e^{i\, 2 g t u}e^{-|t|/2}\sigma(|t|)\, .
\end{equation}
We recall that $u$ is a  rapidity variable originally used to
parameterize the Bethe root distributions of gauge and string
theory Bethe ans\"atze \cite{Beisert:2004hm,Arutyunov:2004vx},
which gives another reason for studying the density of
fluctuations on the $u$-plane. Thus, substituting in eq.(\ref{ft})
the power series expansion for $\sigma(t)$ we get
$$
\sigma(u)=\frac{1}{2\pi}\sum_{n=1}^{\infty}s_n
\int_{-\infty}^\infty{\rm d}t~ e^{i\,  2 g t u}e^{-|t|/2}
\frac{|t|}{e^{|t|}-1}\frac{J_n(2g|t|)}{2g|t|}+\ldots
$$
For large values of $g$ this expression can be well approximated
as
$$
\sigma(u)=\frac{1}{2\pi}\sum_{n=1}^{\infty}s_n
\int_{-\infty}^\infty{\rm d}t~ e^{i\,  2 g t
u}e^{-|t|/2}\frac{J_n(2g|t|)}{2g|t|}+\ldots
$$
The last integral is computed by using the following formula
\cite{Eden:2006rx}
\begin{equation}
\nonumber \int_0^\infty {\rm d}t~ e^{\pm 2 g i u
t}e^{-t/2}\frac{J_n(2gt)}{2gt}=\frac{(2g)^{n-1}}{n}\left(u^{\pm}\left(
1+\sqrt{1+4g^2/(u^\pm)^2} \right) \right)^{-n}\,
\end{equation}
with $u^{\pm}=1/2 \mp 2 i g u$.

\medskip

In this way we obtain the following series representation for the
density $\sigma(u)$ \bea\nonumber \sigma(u)&=&\frac{1}{2\pi
g}\sum_{n=1}^{\infty} s_n f_n+\ldots \, ,\eea where \bea \nonumber
f_n= \frac{(2g)^{n}}{2n}\left[ \big(
u^+(1+\sqrt{1+4g^2/(u^+)^2})\big)^{-n}+\big(
u^-(1+\sqrt{1+4g^2/(u^-)^2}) \big)^{-n} \right] \, .\eea

\medskip

In what follows it is convenient to introduce the expansion
parameter $\epsilon = 1/(2 g)$. Our considerations above suggest
that the density $\sigma(u)$ expands starting from the second
order in $\epsilon$:
\begin{equation}
\label{sdevelopment} \sigma(u) \, = \, \epsilon^2 \sigma^{\ell}(u)
+ \epsilon^3 \sigma^{s\ell}(u) + \ldots
\end{equation}
To find the leading contribution $\sigma^{\ell}(u)$ we have to
develop the large $g$ expansion of the functions $f_n$. The result
is not uniform, it depends on whether $n$ is even or odd and also
on the value of $u$ . For $n$ even we find \bea \nonumber ~~~~~~~
f_{2k}=\left\{
\begin{array}{l} \frac{(-1)^{k}}{2k}T_{2k}(u)+{\cal O}(\epsilon)\,
~~~~~{\rm for}~~~~~|u|<1 \, , \\
\\
\frac{(-1)^{k}}{2k}\Big(u\Big(1+\sqrt{1-\frac{1}{u^2}}\Big)\Big)^{-2k}+{\cal
O}(\epsilon)\, ~~~~~{\rm for}~~~~~|u|>1 \, .
\end{array} \right. \eea
Here $T_{2k}(u)$ are the Chebyshev polynomials of the first kind.
For $n$ odd we obtain  \bea \nonumber
 f_{2k-1}=\left\{
\begin{array}{l} -\frac{(-1)^{k}}{2k-1}\sqrt{1-u^2}~U_{2k-2}(u)+{\cal O}(\epsilon)\,
~~~~~{\rm for}~~~~~|u|<1 \, , \\
\\
~~~0+{\cal O}(\epsilon)\, ~~~~~{\rm for}~~~~~|u|>1 \, .
\end{array} \right. \eea
Here $U_{2k-2}(u)$ are the Chebyshev polynomials of the second
kind. We recall that the Chebyshev polynomials of the first and
the second kind form a sequence of orthogonal polynomials  on the
interval $[-1,1]$ with the weights $(1-u^2)^{-1/2}$ and
$(1-u^2)^{1/2}$ respectively.

\medskip To find the leading density $\sigma^{\ell}(u)$ inside the
interval $|u|<1$ we can use the trigonometric definition of the
Chebyshev polynomials which corresponds to choosing
parametrization $u=\cos\theta$ with $0\leq  \theta\leq \pi$. Thus,
taking the limit $g\to\infty$ we obtain for the leading density
the following expression
\begin{equation}
\sigma^{\ell}(u)=\frac{2}{\pi
}\sum_{k=1}^{\infty}\Big(s_{2k-1}^{\ell}
f_{2k-1}(\theta)+s_{2k}^{\ell} f_{2k}(\theta)\Big)\, ,
\end{equation}
where \bea f_{2k}(\theta)=(-1)^k \frac{\cos 2k \theta }{2k} \, ,
~~~~~~~~ f_{2k-1}(\theta)= -(-1)^k \frac{\sin(2k-1)\theta}{2k-1}\, .
\eea Given the proposal (\ref{sol}) for the coefficients
$s_k^{\ell}$, we can now sum the series for $\sigma^{\ell}(u)$ and
obtain \bea \sigma^{\ell}(u)=\frac{1}{\pi}\, . \eea Thus, inside the
interval $[-1,1]$ the density $\sigma^{\ell}(u)$ is {\it constant}.

\smallskip

\vskip 10pt \noindent
\begin{minipage}{\textwidth}
\begin{center}
\includegraphics[width=0.70\textwidth]{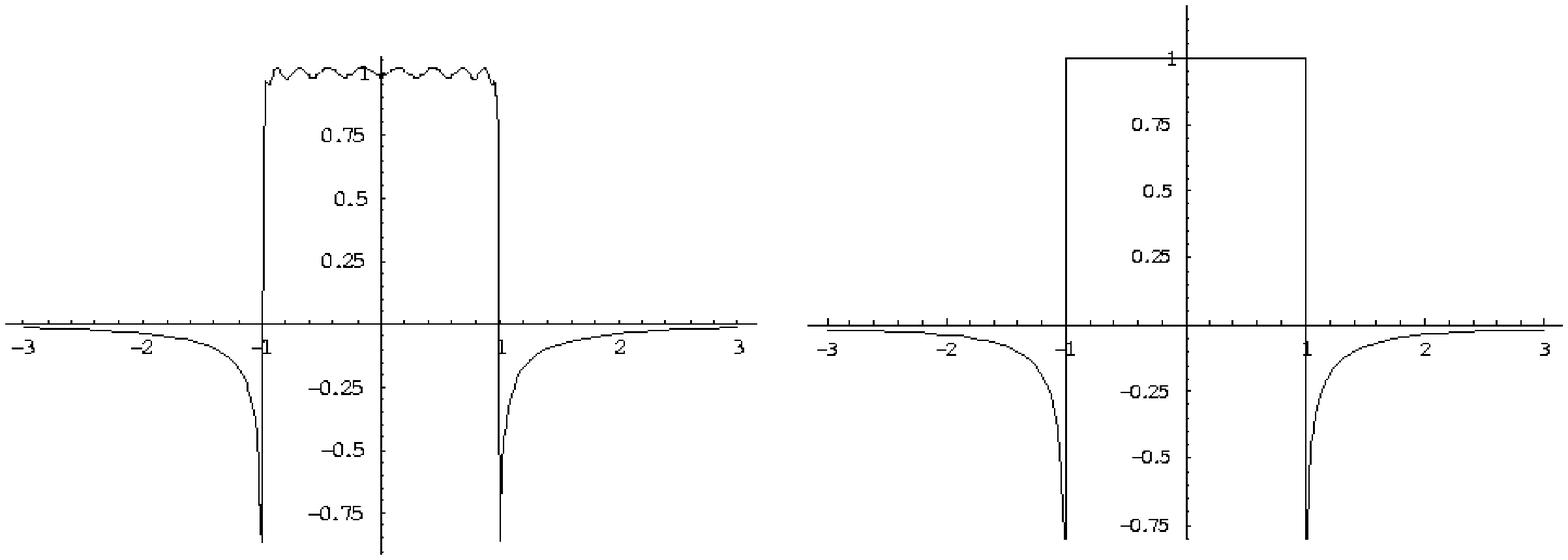}
\end{center}
\end{minipage}
\vskip 15pt
\begin{center}
\parbox{5in}{{\footnotesize Fig.2. The left figure is the plot of a numerical solution for the leading density $\pi\sigma^{\ell}(u) $.
The right figure represents an exact analytic solution for the
same quantity. } }
\end{center}
\vskip 0.3cm

\noindent We note that the functions $f_k(\theta)$ provide a
complete set of functions on the interval $[0, 2 \pi]$, while on
the interval $[0,\pi]$ they are overcomplete. This fact is
consistent with the ambiguity for the coefficients $s_k$ we
observed  in the previous section. Moreover, it is not hard to
convince ourselves that the proposal for $s_k^{\ell}$ discussed in
the previous section is the only possible solution such that
$s_{2k-1}^{\ell}=s_{2k}^{\ell}$ and for which the density
$\sigma^{\ell}(u)$ is constant for $|u|\leq 1$.

\smallskip
Further, it is easy to sum up the series defining the leading
density for $|u|>1$. The result is \bea
\sigma^{\ell}(u)=\frac{1}{2\pi
}\left(2-\frac{\sqrt{2}}{\sqrt{1-u(u\mp \sqrt{u^2-1})}}\right)\, .
\label{tobe} \eea Here the minus and plus signs in the denominator
corresponds to the regions $u>1$ and $u<-1$ respectively. The plot
of the complete analytic solution for $\sigma^{\ell}(u)$ is
presented in Fig.2. It should be compared to the plot of the
numerical solution obtained by using in the series representation
for the density the coefficients $s_k^{\ell}$ obtained from our
numerical analysis. The numerical plot corresponds to taking
$g=40$ and truncating the series at $k=20$.

\smallskip

Finally, we mention the expression for the scaling function $f(g)$
in terms of the density $\sigma(u)$:
$$
f(g)=32g^3\int_{-\infty}^{\infty}{\rm d}u\,
\sigma(u)=8g\int_{-\infty}^{\infty}{\rm d}u\,
\sigma^{\ell}(u)+\ldots=4g+\ldots
$$
This completes our discussion of the leading order analytic
solution of the BES equation in the strong coupling limit.

\subsection{Subleading corrections}
Here we will investigate the first subleading correction to the
leading coefficients $s_k^{\ell}$. By expanding the BES equation we have already obtained equation (\ref{O2o}) which expresses the subleading density $s^{s\ell}_n$ in terms of the leading one:
\begin{equation}\label{sl1}
Z^{\ell}_{2m-1,r} s^{s\ell}_r  = - Z^{s\ell}_{2m-1,r} s^{\ell}_r - {1 \over 2(2m-1)} s^{\ell}_{2m-1}\ .
\end{equation}
Again this equation is degenerate and needs to be supplemented by
a second one that appears at higher order in $1/g$ in the BES
equation. Substituting the explicit form of the coefficients
$s^{\ell}_n$ obtained above, the right hand side of eq.(\ref{sl1})
is seen to vanish (i.e. $K^{(c)\, \ell} s^{s\ell}=0$), which
already implies tight restrictions on the subleading corrections
$s^{s\ell}$. As discussed above, this equation allows for solving
for half of the components. For instance, using eqs.(\ref{evenodd})
and (\ref{inverses}) we can solve for the even components in terms
of the odd ones
\begin{equation}
\label{subconstraint} s^{s\ell}_{2m}=- \sum_{n=1}^\infty
(-1)^{m-n}\frac{32}{\pi}\frac{m^3}{(4m^2-(1-2n)^2)(1-2n)^2}s^{s\ell}_{2n-1}\,
.
\end{equation}

\smallskip
To obtain another equation constraining the subleading solution, we
examine corrections to the BES equation as follows.
%
From the $\mathcal{O}(1/g)$ contribution to the even rows of the BES
equation (\ref{BES2}) we find
\begin{eqnarray} \label{O2e}
s^{\ell}_{2n} &=&8n Z^{ss\ell}_{2n,1} - 4nZ^{\ell}_{2n,m} s^{s\ell}_m - 4nZ^{s\ell}_{2n,m} s^{\ell}_m  \\
&-& 16n(2m-1)\Big[ Z^{\ell}_{2n,2m-1}Z^{\ell}_{2m-1,r} s^{ss\ell}_r + Z^{\ell}_{2n,2m-1}Z^{s\ell}_{2m-1,r} s^{s\ell}_r + Z^{\ell}_{2n,2m-1}Z^{ss\ell}_{2m-1,r} s^{\ell}_r \nonumber \\
&& \qquad\qquad\qquad + Z^{s\ell}_{2n,2m-1}Z^{\ell}_{2m-1,r} s^{s\ell}_r + Z^{s\ell}_{2n,2m-1}Z^{s\ell}_{2m-1,r} s^{\ell}_r + Z^{ss\ell}_{2n,2m-1}Z^{\ell}_{2m-1,r} s^{\ell}_r \Big]\ .\nonumber
\end{eqnarray}
To eliminate the term in $s^{ss\ell}$ we turn to the odd rows of the $\mathcal{O}(1/g^2)$ equation
\begin{equation} \label{O3o}
- 2(2m-1)\Big[Z^{\ell}_{2m-1,r} s^{ss\ell}_r + Z^{s\ell}_{2m-1,r} s^{s\ell}_r + Z^{ss\ell}_{2m-1,r} s^{\ell}_r \Big] = s^{s\ell}_{2m-1}\ .
\end{equation}
Using this together with (\ref{O1o}) and (\ref{O2o}) to simplify (\ref{O2e}) we obtain
\begin{eqnarray}
Z^{\ell}_{2n,m} (-1)^m s^{s\ell}_m = - Z^{s\ell}_{2n,m} (-1)^m s^{\ell}_m - {1 \over 4n}s^{\ell}_{2n}\ .
\end{eqnarray}
This can be combined with (\ref{sl1}) into the full rank equation
\begin{eqnarray}\label{slfullrank}
\tilde{Z}^{\ell}_{nm} s^{s\ell}_m = - \tilde{Z}^{s\ell}_{nm} s^{\ell}_m - {1 \over 2n}s^{\ell}_{n}\ ,
\end{eqnarray}
where again the right hand side vanishes when evaluated on the leading solution $s^{\ell}$ found above.

In view of the divergence of the $\mathcal{O}(1/g)$ correction to
$Z_{nm}$, which could indicate the presence in it of terms e.g.
$\mathcal{O}(\, \log (g)/ g)$, the strong coupling expansion of
$s_n (g)$ may contain more than just pure powers of $1/g$. This
complication begins at $\mathcal{O}(1/g^3)$ in $s_n$ and suggests
that the equations for the subleading terms have to be
supplemented by additional constraints. Luckily, the equations for
the leading term in section 2.3 are not affected by this
complication. We hope to analyze the structure of subleading terms
more fully in future work.
\smallskip

\section{Leading density from the integral equation}
\label{secMe}
In the appendix we compute the inverse Fourier transform of the
BES equation (rather than that of its leading solution) to the
rapidity variable $u$.\footnote{This approach to the BES equation
was also pursued by A. Belitsky (seminar at Princeton University,
December 2006).}  In this section we will expand the BES integral
equation in the small parameter $\epsilon = 1/(2 g)$ assuming that
the density expands as in eq.(\ref{sdevelopment}) and rederive the
leading order solution $\sigma^{\ell}(u)$ by different means.

\smallskip

It turns out that the branch cut of the square root functions
defining the all-loops Bethe-ansatz \cite{Beisert:2005fw} forces
us to distinguish the regimes $|u| \gtrless 1$. The leading order
BES equation constrains the leading density $\sigma^{\ell}(u)$
only within the unit interval; no condition is obtained for
$|u|>1$. We will show that this equation has a unique solution,
namely the constant function $\sigma^{\ell}(u) = 1/\pi, \, |u|<1$.

\smallskip
In the discussion of the last section, the even and the odd
functions, $f_{2k}$ and $f_{2k-1}$, constitute equivalent complete
sets of functions on the interval $|u|\leq 1$. The constraint that
the density be constant in this region allows one to eliminate,
e.g., the coefficients $s^\ell_{2 k-1}$ in favour of $s^\ell_{2
k}$ (or vice versa). Now, to leading order only the even functions
$f_{2k}$ have support outside the interval $|u|< 1$. The knowledge
of the density outside this interval would then fix the remaining
coefficients $s^\ell_{2 k}$.

\smallskip
Above, we have rather taken the opposite approach: pairing the
numerical observation that $s^\ell_{2 k -1} = s^\ell_{2 k}$ with
the requirement that the density be constant within the unit
interval we could equivalently fix the density $\sigma^{\ell}(u)$
on the whole real axis, c.f. eq.(\ref{tobe}).

\smallskip
The leading terms in the strong coupling expansion of the BES
equation are obtained in appendix 5.2. We find
\begin{equation}
\int_{-1}^1 {\rm d}u' \, \sigma^{\ell}(u') \, \hat
K^{(c)\,\ell}(u,u') \, = \, 2 \sqrt{1 - u^2} -
\frac{1}{\sqrt{1-u^2}}  \, , \label{strong1}
\end{equation}
where
\begin{equation}
\hat K^{(c)\,\ell}(u,u') \, = \, - 2 \left(
\frac{\sqrt{1-(u')^2}}{\sqrt{1-u^2}} \, + \, \frac{1}{4} \,\log
\left( \frac{ \left( \sqrt{1-u^2} - \sqrt{1 - (u')^2} \right)^2 }{
\left( \sqrt{1-u^2} + \sqrt{1 - (u')^2} \right)^2 } \right)
\right) \label{kcleading} \, .
\end{equation}
In Fourier space, the potential for the BES equation is actually
the kernel with the second argument put to zero; or to put it
differently, the kernel integrated on a delta function. In the
rapidity plane this means that the potential is the integral of
the kernel on a constant function. By construction, our leading
equation is solved by
\begin{equation}
\sigma^{\ell}(u) \, = \, \frac{1}{\pi} \, , \qquad |u|<1 \, .
\end{equation}
Let us check whether there is a second solution inside the
interval $|u|<1$. To this end, we put
\begin{equation}
U = \sqrt{1-u^2} \, , \qquad V = \sqrt{1-(u')^2}
\end{equation}
and differentiate the whole equation in $U$. Next, we substitute
$U^2 \rightarrow U$ and $V^2 \rightarrow V$ to obtain
\begin{equation}
1 + \frac{1}{2U}\, = \, \int_0^1 {\rm d}V\,  \tilde \sigma(V)
\Big(\frac{1}{U} - \frac{1}{U-V} \Big)\, ,
\end{equation}
where $\tilde \sigma$ is the density written in the new variables
but including the transformation of the integration measure:
\begin{equation}
\tilde \sigma \, = \, \sigma \, \frac{\sqrt{V}}{\sqrt{1-V}}
\end{equation}
There is no regular density that could yield $1/U$ from the
Hilbert transform (i.e. the second term on the r.h.s.). We will
therefore look for a density that produces the constant term on
the l.h.s. from the Hilbert transform, and whose norm is defined
by the $1/U$ terms. Now,
\begin{equation}
\frac{1}{\pi} \int_0^1 {\rm d}V\,  \frac{\sqrt{V}}{\sqrt{1-V}} \,
\frac{1}{U-V} \, = \, - 1 \, , \qquad - \frac{1}{\pi} \int_0^1
{\rm d}V\, \frac{\sqrt{1-V}}{\sqrt{V}} \, \frac{1}{U-V} \, = \, -
1
\end{equation}
and in the first case the normalization is $1/2$, while in the
second it is $-1/2$. Note that the symmetric combination of the
two trial densities
\begin{equation}
\frac{\sqrt{V}}{\sqrt{1-V}} + \frac{\sqrt{1-V}}{\sqrt{V}} \, = \,
\frac{1}{\sqrt{V} \sqrt{1-V}}
\end{equation}
goes to zero under the Hilbert transform whereas it does affect
the norm. Thus, a solution exists and it is unique. We have to
choose
\begin{equation}
\tilde \sigma(V) \, = \, \frac{1}{\pi} \,
\frac{\sqrt{V}}{\sqrt{1-V}}
\end{equation}
which corresponds to $\sigma(u') = 1/\pi$.

\smallskip
In appendix 5.2. we further derive the next-to-leading order of
the BES equation. We find
\begin{equation}
\nonumber \int_{-1}^1 {\rm d}u' \, \sigma^{s\ell}(u') \, \hat
K^{(c)\, \ell}(u,u') \, = \, 0 \, ,~~~~~~:~|u|<1\, .
\end{equation}
By what was said above the leading dressing kernel is invertible
on $[-1,1]$. We therefore conclude
\begin{equation}
\sigma^{s\ell}(u) \, = \, 0 \, , \qquad |u| < 1 .
\end{equation}
No constraint is found on the subleading density outside the unit
interval.

\smallskip
We refrain from pushing the analysis any further for the following
reasons: First, at the next order we would perhaps encounter the
same degeneracy while we lack an independent condition like $s_{2
k}^{\ell} = s_{2k - 1}^{\ell}$ that would allow us to make
progress. Second, the naive expansion in $\epsilon$ quickly leads
to expressions with rather hard singularities in $u,u'$ which are
difficult to handle consistently.
\medskip

The original BES equation has the property that the energy is
given by the value of the density at zero. In the rapidity
variables this means that we can recover the leading contribution
to the energy from the normalization of $\sigma^{\ell}$.
Alternatively, we may use the formula
\begin{eqnarray}
\nonumber E(g) & = & \log(S) \, 8 g^2 \left[1 - 2 g^2
\int_{-\infty}^\infty {\rm d}u \, \sigma(u) \left(
\frac{i}{x^+(u)} - \frac{i}{x^-(u)} \right) \right] \\ & = &
\frac{2 \log(S)}{\epsilon^2} \Biggl[ 1 - \frac{2}{\epsilon^2}
\int_0^1 {\rm d}u \, \sigma(u) \left( 2 \sqrt{1 - u^2} - \epsilon
+ \ldots \right) - \nonumber \\ & & \phantom{\frac{2
\log(s)}{\epsilon^2} \Biggl[ 1 } - \frac{2}{\epsilon^2}
\int_1^\infty {\rm d}u \, \sigma(u) \Biggl(
\frac{\epsilon}{u^2\big(1 + \sqrt{1-\frac{1}{u^2}}\big) - 1} +
\ldots \Biggr) \Biggr]\, . \nonumber
\end{eqnarray}
It is instructive to see how the value of the energy predicted
from string theory is reproduced: The contribution of the constant
part of $\sigma^{\ell}$ within the unit interval cancels the
leading 1 in the square bracket. On the other hand, the
next-to-leading term would receive a contribution from an a priory
non-vanishing subleading density inside the unit interval. We get
\begin{eqnarray}
\nonumber E(g) & = & \log(S)\, 8g\left(\int_0^1{\rm d}u\,
\sigma^{\ell}(u)-\int_1^{\infty}{\rm d}u\,
\frac{\sigma^{\ell}(u)}{u^2\big(1
+ \sqrt{1-\frac{1}{u^2}}\big) - 1}\right.\\
&&~~~~~~~~~~~~~~~~~~~~~~~- \left. 2\int_0^1{\rm d}u\,
\sigma^{s\ell}(u)\sqrt{1-u^2} \right)+\ldots=\log(S) \, 4 g \, +
\ldots \nonumber
\end{eqnarray}
Here we used an identity
$$
\int_0^1{\rm d}u\, \sigma^{\ell}(u)-\int_1^{\infty}{\rm d}u\,
\frac{\sigma^{\ell}(u)}{u^2\big(1 + \sqrt{1-\frac{1}{u^2}}\big) -
1}= \int_{-\infty}^{\infty}{\rm d}u\, \sigma^{\ell}(u)\, .
$$
satisfied by the leading density and also the fact that the
subleading density vanishes inside the unit interval. In fact, the
absence of the subleading density can be considered as the
consistency test on our equations.

\section{Conclusions}
In this paper we analyze the strong coupling limit of the BES
equation which describes the universal scaling function of high
spin operators in ${\cal N}=4$ gauge theory. We have shown that
expanding the BES equation in inverse powers of the coupling constant leads to two equations for the large $g$ solution, one appearing at leading the other at subleading order in $1/g$.
Together they determine a unique
solution in the strong coupling limit whose exact analytic form we present.

\smallskip
Obviously, the next step would be to understand the structure of the
higher order perturbation theory around the strong coupling point
and, in particular, to derive the next-to-leading corrections to the
universal scaling function $f(g)$. We make some progress in this
direction by deriving a pair of equations for the subleading
solution, but also point out some subtleties that arise in
developing the expansion around the leading order solution.

\smallskip

As was shown in \cite{Beisert:2006ez} (see also
\cite{Gomez:2006mf}), the coefficients of the perturbative series
describing the solution of the BES equation at weak coupling admit
 an analytic continuation to strong coupling, where
they coincide with those predicted by string theory. The approach
we adopt here can be considered as another, complementary way to
analytically continue from weak to strong coupling.

\smallskip

On the rapidity $u$-plane the leading fluctuation density
$\sigma^{\ell}(u)$ appears to be constant inside the unit interval
$|u|< 1$. We could argue that this constant part of
$\sigma^{\ell}(u)$ is an artifact of the way the BES equation was
derived: The non-vanishing constant part of the leading density
offsets the splitting of the weak-coupling density into a log
divergent one-loop part and a regular higher loop piece carrying
$\log S$ as a coefficient. Further, as we have shown, the
subleading correction inside the unit interval is absent; in a
manner of speaking a gap opens between $[-\infty,-1]$ and
$[1,\infty]$. This could be qualitatively compared to the results
obtained from string theory. Indeed, the solution of the integral
equation describing the classical spinning strings in ${\rm
AdS}_3\times {\rm S}^1$ \cite{Kazakov:2004nh} in the limit $S
\rightarrow \infty$ with spin $J$ along ${\rm S}^1$ fixed has
support only outside the interval $|u|<1$. The same behavior is
expected\footnote{We would like to thank Sergey Frolov for the
discussion of this point.} for the GKP solution which is obtained
in the limit $J\to 0$. For finite $S$ the solution is elliptic and
it exhibits logarithmic singularities in the limit $S\to\infty$.
On the other hand, our strong coupling density (\ref{tobe}) is an
algebraic function which carries $\log S$ as a normalization. Of
course, this density leads to the same energy as for the GKP
string. Thus, it is desirable to understand the detailed matching
between the string density (higher conserved charges) and the
density we found from the strong coupling limit of the BES
equation. We plan to return to this interesting question
elsewhere.

\section*{Acknowledgements}
We are grateful to Andrei Belitsky, Sergey Frolov and Matthias
Staudacher for valuable discussions. I.~R.~K. and M.~K.~B. would
like to thank Sergio Benvenuti and Antonello Scardicchio for
collaboration on the early stages of this project. This work was
supported in part by the EU-RTN network {\it Constituents,
Fundamental Forces and Symmetries of the Universe}
(MRTN-CT-2004-005104), by the INTAS contract 03-51-6346, by the NWO
grant 047017015, and by the National Science Foundation under grant
No. PHY-0243680. The work of L.~F.~A. was supported by the VENI
grant 680-47-113. The work of G.~A. was supported in part by the
RFBI grant N05-01-00758 and by the grant NSh-672.2006.1.

\section{Appendix}

\subsection{The inverse Fourier transform of the BES equation}
Let us define the kernels
\begin{equation}
K_+(u,u') \, = \, \frac{\left( 1 - \frac{g^2}{x^+(u) \, x^-(u')}
\right) \left( 1 + \frac{g^2}{x^+(u) \, x^-(u')} \right)}{\left( 1
- \frac{g^2}{x^-(u) \, x^+(u')} \right) \left( 1 +
\frac{g^2}{x^-(u) \, x^-(u')} \right)} \, ,
\end{equation}
\begin{equation}
K_-(u,u') \, = \, \frac{\left( 1 + \frac{g^2}{x^-(u) \, x^+(u')}
\right) \left( 1 - \frac{g^2}{x^-(u) \, x^-(u')} \right)}{\left( 1
+ \frac{g^2}{x^+(u) \, x^-(u')} \right) \left( 1 -
\frac{g^2}{x^+(u) \, x^+(u')} \right)} \ ,
\end{equation}
with
\begin{equation}
x^\pm(u) = \frac{1}{2}\left(u \pm \frac{i}{2}\right)\left(1 +
\sqrt{1 -  \frac{4 g^2}{(u \pm \frac{i}{2})^2}} \, \right) \, .
\end{equation}
We further define
\begin{eqnarray}
\hat K_0(u,u') & = & \frac{i}{2} \, \partial_u \log \Bigl(
K_+(u,u') \, K_-(u,u') \Bigr) \, , \label{defK0} \\ \hat K_1(u,u')
& = & \frac{i}{2} \, \partial_u \log \Bigl( K_+(u,u') \, / \,
K_-(u,u') \Bigr) \, . \label{defK1}
\end{eqnarray}
In the same way as in appendix D of \cite{Eden:2006rx} we may show
that
\begin{equation}
\hat K_{0,1}(u,u') \, = \, 2 g^2 \int \int_{-\infty}^\infty {\rm
d}t \, {\rm d}t' e^{i u t + i u' t'} |t| e^{- (|t|+|t'|)/2}
K_{0,1}(2 g |t|, 2 g |t'|) \, ,
\end{equation}
where $K_{0,1}$ are defined in (\ref{defK01}) in the main text.
The dressing kernel is \cite{Beisert:2006ez}
\begin{equation}
K^{(c)}(2 g |t|, 2 g |t'|) \, = \, 2 g^2 \int_{-\infty}^\infty
{\rm d}t'' \, K_1(2 g |t|, 2 g |t''|) \, \frac{|t''|}{e^{|t''|} -
1} \, K_0(2 g |t''|, 2 g |t'|)
\end{equation}
and its Fourier back-transform yields
\begin{eqnarray}
\hat K^{(c)}(u,u') & = & 2 g^2 \int \int_{-\infty}^\infty {\rm d}t
\, {\rm d}t' e^{i u t + i u' t'} |t| e^{- (|t|+|t'|)/2} K^{(c)}(2
g |t|, 2 g |t'|) \nonumber \\ & = & \frac{1}{4 \pi^2} \int \int
\int_{-\infty}^\infty {\rm d}v \, {\rm d}v' \, {\rm d}t'' \, \hat
K_0(u,v) \, \frac{e^{-i (v + v') t''}}{1 - e^{-|t''|}} \, \hat
K_1(v',u') \, . \label{hatKc}
\end{eqnarray}
In the $u,u'$ variables the BES equation becomes:
\begin{eqnarray}
0 & = & 2 \pi \, \sigma(u) \, - \, 2 \int_{-\infty}^\infty {\rm
d}u' \, \sigma(u') \frac{1}{(u-u')^2 + 1} \\ & + &
\int_{-\infty}^\infty {\rm d}u' \, \Big( \sigma(u') - \frac{1}{4
\pi g^2} \Big) \left( \hat K_0(u,u') + \hat K_1(u,u') + 2 \hat
K^{(c)}(u,u') \right) \, .\nonumber
\end{eqnarray}
We have written the potential as the integral of the kernels on
the constant function $-1/(4 \pi g^2)$, i.e. the Fourier
back-transform of $-1/(2 g^2) \, \delta(t')$.

\subsection{The strong coupling limit}

The  integration over $t''$ in eq.(\ref{hatKc}) is not well
defined. To make sense of it we borrow the derivative from $\hat
K_0$ (c.f. (\ref{defK0})) on the right:
\begin{eqnarray}
\partial_{v'} \int_{-\infty}^\infty {\rm d}t'' \frac{e^{- i (v+v') t''}}
{1 - e^{-|t''|}} & = & - i \ \int_{-\infty}^\infty {\rm d}t'' e^{-
i (v+v') t''} \left(\frac{t''}{e^{|t''|} - 1} + t'' \right) \\ & =
&
\partial_{v'} \left( - \Psi(i(v+v')) - \Psi(-i(v+v')) + 2 \pi
\delta(v+v') \right) \, ,\nonumber
\end{eqnarray}
where $\Psi(v)$ is the digamma function. We may then partially
integrate the outer derivative back to act on $\hat K_0$.

\smallskip

The physical situation we consider is in a kinematical regime in
which $u$ scales with $2 g$ at strong coupling, as can be seen for
instance from the numerical studies of the integral equation for
density of fluctuations in \cite{Eden:2006rx}. In the following we
will rescale $u \rightarrow 2 g \, u$ and consider a perturbation
series in $\epsilon = 1/(2 g)$.

\smallskip

In the middle integral in eq.(\ref{hatKc}) we rescale and employ
the asymptotic expansion of the digamma function:
\begin{eqnarray}
&&- \Psi(i(v+v')) - \Psi(-i(v+v')) + 2 \pi \delta(v+v')
\label{logfunc} \\ &  &~~~~~\rightarrow  - \log\left( (v+v')^2
\right) + \frac{\pi}{g} \, \delta(v+v') + {\cal O}(\epsilon^2)
\nonumber
\end{eqnarray}
Note that we have discarded a term $\log(2 g) + c_0$, where $c_0$
is an integration constant from the differentiation trick. A
constant integrated onto the leading order of $\hat K_1$ in the
second argument (or $\hat K_0$ in the first) actually yields zero.
Further, the derivative of the exact expression in terms of the
digamma function is a principal value for $1/(v+v')$ (i.e. a
cut-off at zero). We usually evaluate $\log((v+v')^2)$ under the
subsequent $v,v'$ integrations by partial integration, albeit
using a Cauchy principal value. It ought not matter which
definition is used for the principal value.
\smallskip

After rescaling we find the following leading order dressing
kernel in configuration space:
\begin{equation}
\hat K^{(c)\, \ell}(u, u') \, = \, - \frac{1}{2 \pi^2} \int
\int_{-\infty}^\infty {\rm d}v \, {\rm d}v' \tilde K_1^\ell(u,v)
\left( \frac{1}{2} \log\left( (v+v')^2 \right) \right) \tilde
K_0^\ell(v', u') \, .\label{leadKc}
\end{equation}
Here $\hat K_{0,1} = \epsilon \tilde K_{0,1}$. To leading order
\begin{eqnarray}
& |u| > 1 \; : & x^+(u) \, = \, x^-(u) \, = \, \frac{1}{2
\epsilon} \, u \, \left(1 + \sqrt{1 - \frac{1}{u^2}} \, \right) \,
+ \, \ldots \\ & |u| < 1 \; : & x^+(u) \, = \, \frac{1}{2
\epsilon} \left( u + i \sqrt{1-u^2} \, \right) \, + \, \ldots \\ &
& x^-(u) \, = \, \frac{1}{2 \epsilon} \left( u - i \sqrt{1-u^2} \,
\right)+\ldots  \nonumber
\end{eqnarray}
from which it is easy to obtain
\begin{equation}
\tilde K_0^\ell(u,v) \, = \, 0 \quad : \quad |v| > 1 \, , \qquad
\tilde K_1^\ell(u,v) \, = \, 0 \quad : \quad |u| > 1
\end{equation}
and the non-vanishing sectors
\begin{eqnarray}
& |u| < 1 \; : & \tilde K_0^\ell(u,v) \, = \, \pi (\delta(u-v) +
\delta(u+v)) \, , \\ & |u| > 1 \; : & \tilde K_0^\ell(u,v) \, = \,
- 2 \frac{\sqrt{1-v^2}}{\sqrt{1-\frac{1}{u^2}}} \frac{1}{u^2-v^2}
\, , \nonumber
\end{eqnarray}
and
\begin{eqnarray}
& |v| < 1 \; : & \tilde K_1^\ell(u,v) \, = \, \pi (\delta(u-v) +
\delta(u+v)) - \frac{2}{\sqrt{1-u^2}} \, , \\ & |v| > 1 \; : &
\tilde K_1^\ell(u,v) \, = \, - \frac{2}{\sqrt{1-u^2}} \left( 1 +
\sqrt{1-\frac{1}{v^2}} \, \frac{v^2}{u^2-v^2} \right) \, .
\nonumber
\end{eqnarray}
In particular, the leading contribution (\ref{leadKc}) to $\hat
K^{(c)}(u,u')$ is only non-vanishing when both $u,u'$ lie in the
unit interval. Let us evaluate eq.(\ref{leadKc}) explicitly,
plugging in the formulae for $\tilde K_{0,1}^\ell$ and executing
the two integrations in turn. First,
\begin{eqnarray}
&& \left( \int_{-\infty}^{-1} + \int_1^\infty \right) {\rm d}v' \,
\frac{1}{2} \log \left( (v + v')^2 \right) \, \tilde
K_0^\ell(v',u') \nonumber
\\ & = & - 2 \pi \, \log \left( \sqrt{1-(u')^2} + \sqrt{1 - v^2}
\right) \, \qquad : \qquad |u'|,|v| < 1 \\ & = & - \pi \, \log
\left( v^2 - (u')^2 \right) \, \qquad : \qquad |u'| < 1 \, , \,
|v| > 1\, , \nonumber
\end{eqnarray}
while the $v'$-integration over the unit interval contributes
\begin{equation}
\int_{-1}^1 {\rm d}v' \, \frac{1}{2} \log \left( (v + v')^2
\right) \, \tilde K_0^\ell(v',u') \, = \, \frac{\pi}{2} \log
\left( (v^2 - (u')^2)^2 \right) \, .
\end{equation}
Hence, if $|u'|,|v| < 1$
\begin{eqnarray}
&& \int_{-\infty}^\infty {\rm d}v' \frac{1}{2} \log \left(
(v+v')^2 \right) \, \tilde K_0^\ell(v',u')\\ &\,  =\,  & - 2 \pi
\, \log \left( \sqrt{1-(u')^2} + \sqrt{1 - v^2} \right) +
\frac{\pi}{2} \log \left( (v^2 - (u')^2)^2 \right) \nonumber \\ &
= & \frac{\pi}{2} \log \left( \frac{ \left( \sqrt{1-v^2} - \sqrt{1
- (u')^2} \right)^2 }{ \left( \sqrt{1-v^2} + \sqrt{1 - (u')^2}
\right)^2 } \right) \nonumber
\end{eqnarray}
and otherwise we find zero. It is then easy to see that
\begin{equation}
\hat K^{(c)\, \ell}(u,u') \, = \, - 2 \left(
\frac{\sqrt{1-(u')^2}}{\sqrt{1-u^2}} \, + \, \frac{1}{4} \,\log
\left( \frac{ \left( \sqrt{1-u^2} - \sqrt{1 - (u')^2} \right)^2 }{
\left( \sqrt{1-u^2} + \sqrt{1 - (u')^2} \right)^2 } \right)
\right) \nonumber
\end{equation}
to leading order  in the strong coupling expansion.
\medskip

The rescaled BES equation is
\begin{eqnarray}
0 & = & 2 \pi \, \sigma(u) \, - \, 2 \int_{-\infty}^\infty {\rm
d}u' \, \sigma(u') \frac{\epsilon}{(u-u')^2 + \epsilon^2} \\ & + &
\int_{-\infty}^\infty {\rm d}u' \, \left( \sigma(u') -
\frac{\epsilon^2}{\pi} \right) \left( \tilde K_0(u,u') + \tilde
K_1(u,u') + \frac{2}{\epsilon} \hat K^{(c)}(u,u') \right) \, .
\nonumber
\end{eqnarray}
Note that the kernel in the second term on the r.h.s. is a
representation of $\pi \, \delta(u-u')$ so that the first two
terms cancel for small $\epsilon$. The numerical analysis
displayed  in the paper suggests to look for a solution
$$\sigma(u') = \epsilon^2 \, \sigma^{\ell}(u') \, + \, \epsilon^3 \,
\sigma^{s\ell}(u') \, + \, {\cal O}(\epsilon^4).$$ The leading
equation is
\begin{equation}
\int_{-1}^1 {\rm d}u' \left( \sigma^{\ell}(u') - \frac{1}{\pi}
\right) \hat K^{(c)\, \ell}(u,u') = 0 \label{eqlead1} \, .
\end{equation}
It constrains the density $\sigma^{\ell}$ only inside the unit
interval. We prove in the main text that the obvious solution
\begin{equation}
\sigma^{\ell}(u') \, = \, \frac{1}{\pi} \qquad : \qquad |u'| < 1
\end{equation}
is also the only one. At the next order we find the equation
\begin{eqnarray}
0 & = & \left(\int_{-\infty}^{-1} + \int_1^\infty \right) {\rm
d}u' \, \left( \sigma^{\ell}(u') - \frac{1}{\pi} \right) \left(
\tilde K_0^\ell(u,u') + \tilde K_1^\ell(u,u') + 2 \hat K^{(c)\,
s\ell}(u,u') \right) \nonumber \\ & + & \int_{-1}^1 {\rm d}u' \,
\sigma^{s\ell}(u') \hat K^{(c)\, \ell}(u,u')\, . \label{subBES}
\end{eqnarray}
Now, to leading order $\tilde K_0(u,u')$ vanishes when $|u'|>1$ so
that the $\tilde K_0^\ell$ term can be dropped. We show below that
the other two kernels in the first line cancel. Hence for any
$\sigma^\ell$ the last equation reduces to
\begin{equation}
\nonumber \int_{-1}^1 {\rm d}u' \, \sigma^{s\ell}(u') \, \hat
K^{(c)\, \ell}(u,u') \, = \, 0 \, .
\end{equation}
It follows that
\begin{equation}
\sigma^{s\ell}(u') \, = \, 0 \qquad : \qquad |u'| < 1
\end{equation}
because $\hat K^{(c)\, s\ell}$ is invertible on the unit interval
(see Section \ref{secMe}). At this stage we cannot make a
statement about $\sigma^{s\ell}$ outside the unit interval.

\smallskip

To show the aforementioned cancellation of the two kernels, let us
work out $\hat K^{(c) \, s\ell}(u,u')$ for $|u'|>1$. In this case,
$\tilde K_0^\ell(v',u')$ in (\ref{leadKc}) vanishes, so that we
pick up the subleading correction in the right factor while the
other two terms may still be taken at leading order. From eq.
(\ref{defK0}) we can work out
\begin{eqnarray}
& |u| < 1, \, |u'| > 1 \; : & \tilde K_0^{s \ell}(u,u') \, = \,
 \, \frac{u^2 + (u')^2}{(u^2-(u')^2)^2} \, , \\ \nonumber
& |u| > 1, \, |u'|
> 1 \; : & \tilde K_0^{s \ell}(u,u') \, = \,
\frac{1-u^2-(u')^2}{\Delta} \, ,
\end{eqnarray}
where we have introduced the notation
\begin{eqnarray}
\Delta & = & (u^2 (u')^2 + |u u'| \sqrt{u^2-1} \sqrt{(u')^2-1})
(-2 + u^2 + (u')^2) \nonumber \\ & & - u^2(u^2-1) - (u')^2
((u')^2-1) \, .\nonumber
\end{eqnarray}
It is not too hard to check that
\begin{eqnarray}
&& \int_{-\infty}^\infty \frac{1}{2} \log( (v+v')^2) \, \tilde
K_0^{s \ell}(v',u') \nonumber \\ & = & - \frac{\pi}{(u')^2 - v^2}
\frac{\sqrt{1-v^2}}{\sqrt{1-\frac{1}{(u')^2}}} \qquad : \qquad |v|
< 1, \, |u'| > 1 \, , \\
& = & 0 \qquad : \qquad |v|,|u'| > 1  \nonumber
\end{eqnarray}
from which we may easily deduce that
\begin{equation}
\hat K^{(c) \, s \ell}(u,u') \, = \, - \frac{1}{2} \, \tilde
K_1^\ell(u,u') \qquad : \qquad |u| < 1, \, |u'|>1 \, .
\end{equation}


\end{document}